\documentclass[twocolumn,nofootinbib,prd,showpacs,aps,epsf,amsfonts,amssymb,amsbsy]{revtex4}

\usepackage{graphics,bm}
\usepackage{epsfig}
\usepackage{graphicx}

\newcommand{\beq}{\begin{equation}}
\newcommand{\eeq}{\end{equation}}
\newcommand{\bqa}{\begin{eqnarray}}
\newcommand{\eqa}{\end{eqnarray}}

\def\sumint{\hbox{$\sum$}\!\!\!\!\!\!\int}
\def\square{\vcenter{\vbox{\hrule height.4pt
          \hbox{\vrule width.4pt height4pt
          \kern4pt\vrule width.3pt}\hrule height.4pt}}}

\begin{document}

\author{Jens O. Andersen} 
\email{andersen@tf.phys.no}
\author{Rashid Khan}
\email{rashid.khan@ntnu.no}
\affiliation{Department of Physics, 
Norwegian Institute of Science and Technology, N-7491 Trondheim, Norway}
\title{Chiral transition in a magnetic field and at finite baryon density}

\date{\today}

\begin{abstract}
We consider the quark-meson model with two quark flavors
in a constant external magnetic field $B$ at finite
temperature $T$ and finite baryon chemical potential  $\mu_B$.
We calculate the full renormalized effective potential to one-loop
order in perturbation theory. 
We study the system in the large-$N_c$ limit, 
where we treat the bosonic modes at tree 
level. It is shown that the system exhibits 
dynamical chiral symmetry breaking, i. e. 
that an arbitrarily weak magnetic field breaks 
chiral symmetry dynamically, in agreement with earlier calculations
using the NJL model.
We study the influence on the phase transition
of the fermionic vacuum fluctuations.
For strong magnetic fields, $|qB|\sim5m_{\pi}^2$
and in the chiral limit, the transition is first
order in the entire $\mu_B-T$ plane if vacuum fluctuations are not included
and second order if they are included.
At the physical point, the transition is a crossover 
for $\mu_B=0$ with and without vacuum fluctuations.

\end{abstract}

\maketitle


\section{Introduction}
A tremendous amount of work has been done in recent years 
to map out the QCD phase diagram as a function of temperature $T$ and
baryon chemical potential $\mu_B$. 
Two aspects have received particular attention: the 
position of the critical endpoint where the curve of
first-order chiral transitions terminates in a second-order 
transition~\cite{stephanov}, and the
various color-superconducting phases at large baryon chemical potential and
low temperature~\cite{schmitt}.

With $N_f$ quark flavors, the global symmetry
of QCD is $SU(N_f)_V\times SU(N_f)_A$ in the chiral limit
and $SU(N_f)_V$ if the quark masses are equal.
In the chiral limit with $N_f=2$, we 
use the isomorphism between the groups $SU(2)\times SU(2)$ and $O(4)$
and use the $O(4)$ linear sigma model (LSM) as a low-energy
effective theory for QCD.
Coupling the linear sigma model to quarks we obtain the 
quark-meson (QM) model.
Since we have quark degrees of freedom, we can couple the
model to a baryon chemical potential $\mu_B$ and study finite-density effects.
The QM model has been used to study various aspects of the chiral
transition at $\mu_B=0$~\cite{caldas,Scavenius:2000qd,mocsy,Bowman:2008kc,bj} 
and
$\mu_B\neq0$~\cite{erg,Kovacs:2006ym,khan}. 
Schwinger-Dyson equations were used in~\cite{fischer}.
One can couple the Polyakov loop to the quark sector 
by introducing a constant temporal gauge field background~\cite{fukushima}, 
in order to incorporate certain aspects
of the deconfinement transition.
The thermodynamics of the Polyakov-extended quark-meson model 
(PQM) was studied in 
Refs.~\cite{pnjlrat,poly1,nakano,tuo,skokov2,skokov3,marko,polypaw,indiapol},
and the PNJL model in Refs.~\cite{tuo,abuki,weise1,weise2,bl0,bl1,sakai,xuhuang}

Using the QM model, it is a common approximation to omit the
quantum and thermal fluctuations of the bosonic degrees of freedom, i.e.
treating them at tree level. This is the large-$N_c$ limit.
On the other hand, one keeps the 
thermal fluctuations of the quarks. 
Since symmetry breaking takes place in the mesonic sector and at
tree level, the vacuum 
fluctuations of the quarks are sometimes omitted; however one must be
careful when making such additional approximations.
It was shown in Ref.~\cite{bj} that the order of the phase transition
depends on whether or not one includes the fermionic vacuum contributions
in effective potential. This issue was studied in detail in~\cite{khan}
using optimized perturbation theory~\cite{chiku}. 
If all bosonic and fermionic vacuum and thermal
contributions are included, resummed one-loop results predict a
first-order transition in the entire $\mu$--$T$ plane in the chiral limit.

For $N_f=2$ and in the chiral limit, universality and renormalization arguments 
firmly establish the second-order nature of the transition at zero
baryon chemical potential if the axial $U(1)$ symmetry is explicitly broken.
Otherwise it is driven first order.
For $N_f=3$, the transition is always first order in the limit of 
zero current quark masses.
~\cite{pisarski,stephanov}.
At the physical point, i. e. for current quark masses that yield
$m_{\pi}=140$ MeV, it is a crossover transition. At nonzero baryon chemical
potential $\mu_B$ and in particular at $T=0$, the order of the chiral transition
is not obvious from universality 
arguments~\cite{qcdmu}. 
Most model calculations predict a first-order
transition at $T=0$~\cite{stephanov,bub}, although the results depend
on the input parameters and are sensitive to the ultraviolet cutoff.
If there is a first-order transition at $T=0$, there is a line
of first order transitions in the $\mu_B$-$T$ plane
that ends at critical point since we known that the
transition is in a second order for $\mu_B=0$.

Another important question is the behavior of QCD in external magnetic 
fields.
This is not a purely academic question since
the properties of QCD in strong magnetic fields $B$ is relevant in several 
situations. For example, 
large magnetic fields exist inside ordinary
neutron stars as well as magnetars~\cite{neutron}. 
In the latter case, the cores
may be color superconducting and so it is important to study the effects
of external magnetic fields in this phase.
Similarly, it has been suggested that strong magnetic fields are created
in heavy-ion collisions at the Relativistic Heavy-Ion Collider (RHIC) and 
the Large Hadron Collider (LHC) and that these play an important 
role~\cite{harmen1}.
In this case, the magnetic field  strength 
has been estimated to be up to $B\sim 10^{19}$ Gauss, which
corresponds to $|qB|\sim6m_{\pi}^2$, where $|q|$ is the charge of the pion.
One of the interesting issues that has been raised is whether 
there is a splitting of the deconfinement and chiral transitions 
at finite $B$.
This has spurred the interest in studying QCD in external fields.
At zero baryon chemical potential this can be done from first principles
using lattice simulations~\cite{sanf,negro, budaleik}.
At finite $\mu_B$ this is very difficult
due to the infamous sign problem. Therefore one often resorts to
effective theories that share some of the features of QCD, such as 
chiral symmetry breaking. 

A low-energy effective theory that provides a systematic
framework for systematic calculations is chiral perturbation theory.
Chiral perturbation theory has been used to study the
quark condensate in strong magnetic fields at zero 
temperature~\cite{smilga,cptB}, while the quark-hadron phase transition
was investigated in Ref.~\cite{chiralB}.
The effects of external magnetic fields 
on the chiral transition have been studied in detail
using the NJL model~\cite{klev,shovkovy+,gorbie,klim,hiller,boomsma2,chat,avan,frasca,rabbi},
the Polyakov-loop extended NJL model~\cite{pnjlgat,pnjlkas,mfc},
the QM model~\cite{fraga1,frasca,rabbi}, the (P)QM model~\cite{fragapol},
and the instanton-liquid model~\cite{korea}.
Similarly,
the effects of strong magnetic field on the various color superconducting
phases have been studied using the NJL 
model~\cite{qcdmag1,qcdmag2,fh,jorge,qcdmag3}.

In this paper, we  study the effects of magnetic fields 
on the chiral phase transition at finite temperature and finite baryon density.
In order to do so, we employ the quark-meson model with two flavors
coupled to a baryon chemical potential $\mu_B$.
We consider the large $N_c$-limit which amounts to treat the sigma and
pions at tree level. In this limit, we investigate the role of the
fermionic vacuum fluctuations.


The paper is organized as follows.
In Sec.~II, we briefly discuss the QM model in an external magnetic field
and at finite $\mu_B$. In Sec.~III, we calculate the 
standard one-loop effective
potential. In Sec.~IV we discuss the thermodynamics and the numerical
results in the large-$N_c$ limit. In Sec.~V, we summarize and conclude.

\section{Quark-Meson model}
The Euclidean Lagrangian for the quark-meson model with $N_f=2$
flavors is
\bqa
{\cal L}&=&
{\cal L}_{\rm meson}+{\cal L}_{\rm quark}+{\cal L}_{\rm Yukawa}
+{\cal L}_{\rm det}\;,
\label{lagra}
\eqa
where the various terms are
\bqa\nonumber
{\cal L}_{\rm meson}&=&
{\rm Tr}\left[
\partial_{\mu}\Phi^{\dagger}\partial_{\mu}\Phi\right]
+
m^2{\rm Tr}\left[\Phi^{\dagger}
\Phi\right]
\\&&
+{\lambda\over3}{\rm Tr}\left[\Phi^{\dagger}\Phi\right]^2
-{1\over2}h{\rm Tr}
\left[\Phi+\Phi^{\dagger}\right]\;, \\
{\cal L}_{\rm quark}&=&
\bar{\psi}\left[
\gamma_{\mu}\partial_{\mu}-\mu\gamma_4\right]\psi\;, \\
{\cal L}_{\rm Yukawa}&=&
g\bar{\psi}\left[\sigma-i\gamma_5{\boldsymbol \tau}\cdot{\boldsymbol\pi}
\right]\psi\;,
\\ 
\label{det}
{\cal L}_{\rm det}&=&
c\det[\Phi+\Phi^{\dagger}]\;,
\eqa
where
\bqa
\Phi&=&{1\over2}\left(\sigma+
{\boldsymbol \tau}\cdot{\boldsymbol\pi}\right)\;.
\eqa
Here, $\sigma$ is the sigma field, ${\boldsymbol \pi}$ denotes the 
neutral and charged pions. Moreover
${\boldsymbol \tau}$ are the Pauli matrices,
$\mu=\mbox{$1\over2$}(\mu_u+\mu_d)$ 
is the quark chemical potential, where
$\mu_u$ and $\mu_d$ are the chemical potential for the $u$ and $d$ quarks,
respectively. The baryon chemical potential is given by $\mu_B=3\mu$. 
We set $\mu_u=\mu_d$ so that we are working at zero isospin
chemical potential, $\mu_I=\mbox{$1\over2$}(\mu_u-\mu_d)=0$.
The Euclidean $\gamma$ matrices
are given by $\gamma_j=i\gamma^j_M$ and $\gamma_4=\gamma^0_M$, where
the index $M$ denotes Minkowski space.
The fermion field is an isospin doublet
\bqa
\psi=
\left(\begin{array}{c}
u\\
d\\
\end{array}\right)\;.
\label{d0}
\eqa
If $h=0$, the first three terms in Eq.~(\ref{lagra}) are invariant under
$U(2)_L\times U(2)_R\sim SU(2)_L\times SU(2)_R\times U(1)_B\times U(1)_A$.
If $h\neq0$, chiral symmetry is explicitly broken, otherwise
it is spontaneously broken in the vacuum. Either way, the symmetry is
reduced to  
$SU(2)_V\times U(1)_B\times U(1)_A$.
Note that this requires $m^2<0$ which is assumed in the
remainder of the paper. 
The $U(1)_A$ symmetry is also broken in the vacuum
by instantons~\cite{tuft}, and their effects are mimicked by the 
the determinant term Eq.~(\ref{det}). Since the $U(1)_B$ symmetry
is always respected,
symmetry of the QCD  vacuum  is $SU(2)_V\times U(1)_B$.
In the following we set $c=0$ for simplicity.

Chiral symmetry is broken in the vacuum by a nonzero expectation value 
$v$
for the sigma field. We therefore make the replacement
\bqa
\sigma&\rightarrow&v+\tilde{\sigma}\;,
\label{shift}
\eqa
where $\tilde{\sigma}$ is a quantum fluctuating field with vanishing expecation
value. After the shift~(\ref{shift}), 
the tree-level potential is given by
\bqa
{\cal V}_{\rm tree}
&=&{1\over2}m^2v^2+{\lambda\over24}v^4
-hv+{1\over2}B^2+{\cal E}_0
\;,
\label{treeb}
\eqa
where the last term is the vacuum energy density .
The tree-level masses for the sigma and neutral pion are
\bqa
m_{\sigma}^2&=&m^2+{\lambda\over2} v^2\;,\\
m_{\pi}^2&=&m^2+{\lambda\over6} v^2\;.
\eqa
Note that the pion mass $m_{\pi}^2$ vanishes at the minimum of the 
tree-level potential in accordance with Goldstone's theorem.
The charged pions $\pi^{\pm}$ and the quarks
couple to the external magnetic field.
This coupling is implemented by the substitution 
$\partial_{\mu}\rightarrow\partial_{\mu}+iqA_{\mu}$, where
$A_{\mu}$ is the four-vector potential and $q$ is the electric charge
of the particle. 
For a constant magnetic field 
${B}$ in the
$z$-direction, one can conveniently choose the four-vector potential as
$(A_0,{\bf A})=(0,0,Bx,0)$.
The classical solutions to the Klein-Gordon equation in a constant magnetic
field are well known and the dispersion relation is given by
\bqa
\left(E_{n,p_z}^{\pm}\right)^2&=&p_z^2+m^2+{1\over6}\lambda v^2+(2n+1)|qB|\;,
\eqa
where 
$n$ is an nonnegative integer, $q$ is the electric charge of the pion,
and $p_z$ is the spatial momentum
in the $z$-direction. The subscript $\pm$ denotes $\pi^{\pm}$ 
and we note that the dispersion relations are identical.
Similarly, the Dirac equation in a constant magnetic field $B$
can be solved
straightforwardly and the dispersion relation for the quarks is given by
\bqa
E_{n,p_z}^2&=&p_z^2+m_{q}^2+(2n+1-s)|q_fB|\;,
\eqa
where $m_q=gv$ is the quark mass after symmetry breaking, 
$q_f$ is the electric charge of the quark,
and 
$s=\pm1$ denote spin up/down, respectively.

\section{One-loop effective potential}
In this section, we calculate the one-loop effective potential of the 
quark-meson model. This is done by taking into account the Gaussian fluctuations
around the mean field $v$. 
The one-loop contribution ${\cal V}_1$ to the effective potential
can be written as
the sum of the contributions from the $\sigma$, pions, and quarks. 
This yields
\bqa
{\cal V}_1&=&
{\cal V}_{\sigma}+{\cal V}_{\pi^0}
+{\cal V}_{\pi^{+}}+{\cal V}_{\pi^{-}}
+{\cal V}_{\rm q}\;,
\eqa
where
\bqa
\label{sigmi}
{\cal V}_{\sigma}&=&{1\over2}
\sumint_P\log\left[{P^2+m_{\sigma}^2}\right]\;,\\
{\cal V}_{\pi^0}&=&{1\over2}\sumint_P\log\left[{P^2+m_{\pi}^2}\right]\;, \\
{\cal V}_{\pi^{\pm}}&=& 
{1\over2}
{|qB|T\over2\pi}
\sum_{P_0,n}
\int_{p_z}\log\left[P_0^2+p_z^2+M^2_B\right]\;, 
\\ \nonumber
{\cal V}_{\rm q}&=&
-\sum_{f}
{\rm Tr}\log\left[
i\gamma_{\mu}(P_{\mu}+q_fA_{\mu})+m_q-\mu\gamma_4
\right]\;, 
\\ &&
\label{fpisum} 
\eqa
where $M^2_B=m_{\pi}^2+(2n+1)|qB|$ and $f$ denotes the flavor.
The symbol $\sumint_P$ is short-hand notation for
\bqa
\sumint_P&=&
\left({e^{\gamma_E}\Lambda^2\over4\pi}\right)^{\epsilon}
T\sum_{P_0=2\pi nT}\int_p{d^dp\over(2\pi)^d}\;,
\eqa
where $d=3-2\epsilon$ and $\Lambda$ is the renormalization scale associated 
with dimensional regularization in the $\overline{\rm MS}$ scheme.
The integral over spatial momenta will be calculated in dimensional 
regularization.
In the case of the charged pions,
the sum-integral is replaced by a sum over Matsubara 
frequencies $P_0=2\pi nT$,
a sum over Landau levels $n$, and an integral over momenta in $d-2=1-2\epsilon$
dimensions:
\bqa
\sumint_P&\rightarrow&
{|qB|T\over2\pi}
\left({e^{\gamma_E}\Lambda^2\over4\pi}\right)^{\epsilon}\sum_{P_0,n}
\int_{p_z}\;,
\eqa
where the prefactor ${|qB|\over2\pi}$ takes into account the degeneracy
of the Landau levels and 
\bqa
\int_{p_z}&=&\int{d^{d-2}p_z\over(2\pi)^{d-2}}\;,
\eqa 
Similarly, the contribution from the quarks can be written as
\bqa\nonumber
{\cal V}_q&=&
-N_c\sum_{\{P_0\},n,s,f}{|q_fB|\over2\pi}\int_{p_z}
\log\left[P_0^2+p_z^2+M_q^2\right]\;.
\\ &&
\eqa
where $M_q^2=m_{q}^2+(2n+1-s)|q_fB|$, $N_c$ is the number of colors, and
the Matsubara frequencies are $P_0=(2n+1)\pi T$.

The sum-integrals involving the contribution from 
the sigma and the neutral pion are essentially the same
so we consider Eq.~(\ref{sigmi}) with a general mass $M$.
Summing over Matsubara frequencies we can write
\bqa\nonumber
\sumint_P\log\left[{P^2+M^2}\right]
&=&\int_p\bigg\{\sqrt{p^2+M^2} 
\\&&
\hspace{-1cm}
+2T\log\left[1-e^{-\beta\sqrt{p^2+M^2}}
\right]\bigg\}
\;.
\label{divi}
\eqa
The first term in Eq.~(\ref{divi}) 
is ultraviolet divergent. Calculating it with dimensional 
regularization and expanding in powers of $\epsilon$
through order $\epsilon^0$, we obtain
\bqa\nonumber
\sumint_P\log\left[{P^2+M^2}\right]
&=&-{M^4\over32\pi^2}\left({\Lambda^2\over M^2}\right)^{\epsilon}
\left[
{1\over\epsilon}+{3\over2}
\right]
\\&&
\hspace{-1.7cm}
+{T\over\pi^2}\int dp\,p^2\log\left[1-e^{-\beta\sqrt{p^2+M^2}}\right]\;.
\label{sigdiv}
\eqa
We next consider the 
contribution to the effective potential from the charged pions.
Summing over the Matsubara frequencies  in Eq.~(\ref{fpisum}), we obtain
\bqa\nonumber
{\cal V}_{\pi^{\pm}}&=&
{|qB|\over4\pi}\sum_n
\int_{p_z}\left\{
\sqrt{p_z^2+M^2_B}
\right.
\\ &&\left.\hspace{-1cm}
+2T\log\left[1-e^{-\beta\sqrt{p^2+M^2_B}}\right]\right\}
\;.
\eqa
The first integral is ultraviolet divergent and we compute in dimensional
regularization with $d=1-2\epsilon$. This yields
\bqa
\int_{p_z}\sqrt{p_z^2+M^2_B}
&=&
-{M^2_B\over4\pi}
\left({e^{\gamma_E}
\Lambda^2\over M^2_B}\right)^{\epsilon}
\Gamma(-1+\epsilon)\;.
\label{10e}
\eqa
Eq.~(\ref{10e}) shows that the sum over Landau levels $n$
involves the term $M_B^{2-2\epsilon}$. This sum is divergent for $\epsilon=0$
and we regulate it using zeta-function regularization.
After scaling out a factor
of $(2|qB|)^{1-\epsilon}$, this sum can be written as
\bqa\nonumber
\sum_nM_B^{2-2\epsilon}&=&
(2|qB|)^{1-\epsilon}
\sum_n\left[n+\mbox{$1\over2$}
+{m_{\pi}^2\over2|qB|}
\right]^{1-\epsilon}\\
&=&
(2|qB|)^{1-\epsilon}
\zeta(-1+\epsilon,\mbox{$1\over2$}+x)\;,
\eqa
where $x={m_{\pi}^2\over2|qB|}$ and $\zeta(q,s)$ is the Hurwitz zeta function.
The vacuum contribution then reduces to
\bqa\nonumber
{\cal V}_{\pi^{\pm}}^{\rm vac}
&=&-{(qB)^2\over8\pi^2}
\left({e^{\gamma_E}\Lambda^2\over2|qB|}\right)^{\epsilon}
\Gamma(-1+\epsilon)
\zeta(-1+\epsilon,\mbox{$1\over2$}+x)\;.
\\ &&
\label{vakleik}
\eqa
\begin{widetext}
Expanding Eq.~(\ref{vakleik}), we obtain
\bqa
{\cal V}_{\pi^{\pm}}^{\rm vac}&=&
{1\over64\pi^2}
\left({\Lambda^2\over2|qB|}\right)^{\epsilon}
\left[
\left({(qB)^2\over3}-
m^4_{\pi}\right)\left({1\over\epsilon}+1\right)
+8(qB)^2\zeta^{(1,0)}(-1,\mbox{$1\over2$}+x)
+{\cal O}(\epsilon)
\right]\;,
\label{pidiv}
\eqa
where $\zeta^{(1,0)}(-1,{1\over2}+x)$ is the derivative 
of the 
Hurwitz zeta function
with respect to the first argument and where we have used that
$\zeta(-1,{1\over2}+x)={1\over24}-{1\over2}x^2$.

The vacuum contributions from the quarks can be calculated in the same manner,
and one finds~\cite{pinto,ebert+,boomsma2}
\bqa\nonumber
{\cal V}_{q}^{\rm vac}&=&
{N_c\over2\pi^2}\sum_f(q_fB)^2
\left({e^{\gamma_E}\Lambda^2\over2|q_fB|}\right)^{\epsilon}
\Gamma(-1+\epsilon)
\left[
\zeta(-1+\epsilon,x_f)-{1\over2}x_f^{1-\epsilon}
\right]\\ \nonumber
&=&
{N_c\over16\pi^2}
\sum_f
\left({\Lambda^2\over2|q_fB|}\right)^{\epsilon}
\left[
\left({2(q_fB)^2\over3}+m^4_{q}\right)\left({1\over\epsilon}+1\right)
-8(q_fB)^2\zeta^{(1,0)}(-1,x_f)
-2|q_fB|m_q^2\log x_f
+{\cal O}(\epsilon)
\right]\;,\\
&&
\label{qdiv}
\eqa
\end{widetext}
where $x_f=m_q^2/2|q_fB|$.
The divergences of the effective potential are given by 
Eqs.~(\ref{sigdiv}),~(\ref{pidiv}), and~(\ref{qdiv}). 
The divergences that depend on the magnetic field are given by
\bqa
{\cal V}_1^{\rm div}&=&
{(qB)^2\over96\pi^2\epsilon}
+N_c\sum_f{(q_fB)^2\over24\pi^2\epsilon}\;.
\eqa
These divergences are removed by wavefunction renormalization
of the (external) gauge field $A_{\mu}$.
This is done by making the replacement in the tree-level Lagrangian 
Eq.~(\ref{treeb})~\cite{elm}:
\bqa
B^2\rightarrow B^2\bigg[1-{q^2\over48\pi^2\epsilon}
-N_c\sum_f{q^2_f\over12\pi^2\epsilon}
\bigg]\;.
\eqa
The remaining divergences in ${\cal V}_1$ are given by
\bqa\nonumber
{\cal V}_1^{\rm div}&=&
-{1\over64\pi^2\epsilon}\left[
m_{\sigma}^4+3m_{\pi}^4
-4N_cN_fm_{q}^4\right]\;.
\eqa
These are the same divergences as one encounters in vanishing magnetic field
and so the usual renormalization procedure can be used to eliminate them.
This is done by adding a vacuum energy counterterm $\Delta{\cal E}_0$
and making
the replacements 
$m^2\rightarrow m^2+\delta m^2$
and $\lambda\rightarrow\lambda+\delta\lambda$
in the tree-level effective 
potential~(\ref{treeb}), where~
\bqa
\Delta{\cal E}_0&=&{m^4\over16\pi^2\epsilon}\;,\\
\delta m^2&=&{\lambda m^2\over16\pi^2\epsilon}\;,\\
\delta\lambda&=&
{\lambda^2\over8\pi^2\epsilon}
-{3N_cN_fg^4\over2\pi^2\epsilon}
\;.
\eqa
The renormalized one-loop contribution to the
effective potential at $T=0$ then becomes
\begin{widetext}
\bqa\nonumber
{\cal V}_{1}^{\rm vac}&=&-
{m_{\sigma}^4\over64\pi^2}\left[
\log{\Lambda^2\over m_{\sigma}^2}+{3\over2}\right]-
{m_{\pi}^4\over64\pi^2}\left[
\log{\Lambda^2\over m_{\pi}^2}+{3\over2}\right]
-{m_{\pi}^4\over32\pi^2}\left[
\log{\Lambda^2\over2|qB|}+1
\right]
+
{N_cm_{q}^4\over16\pi^2}
\sum_{f}
\left[
\log{\Lambda^2\over2|q_fB|}+1
\right]
\\&& \nonumber
+{(qB)^2\over4\pi^2}\zeta^{(1,0)}(-1,{1\over2}+x)
-{N_c\over2\pi^2}\sum_f
(q_fB)^2
\zeta^{(1,0)}(-1,x_f)
-{N_cm_{q}^2\over8\pi^2}\sum_f|q_fB|\log{m_{q}^2\over2|q_fB|}
\\ &&
+{(qB)^2\over96\pi^2}\log{\Lambda^2\over2|qB|}
+N_c\sum_f{(q_fB)^2\over24\pi^2}\log{\Lambda^2\over2|q_fB|}
\;.
\label{finvac}
\eqa

The finite-temperature contribution ${\cal V}_1^{T,\mu}$ 
to the effective potential is given by
\bqa\nonumber
{\cal V}_{1}^{T,\mu}
&=& \nonumber{T\over2\pi^2}\int_0^{\infty}
dp\,p^2\log\left[1-e^{-\beta\sqrt{p^2+m_{\sigma}^2}}\right]
+{T\over2\pi^2}\int_0^{\infty}
dp\,p^2\log\left[1-e^{-\beta\sqrt{p^2+m_{\pi}^2}}\right]
\\ && 
+
{|qB|T\over\pi^2}\sum_n\int_0^{\infty}
dp\,\log\left[1-e^{-\beta\sqrt{p^2+M_B^2}}\right]
-N_c\sum_{s,f,n}{|q_fB|T\over2\pi^2}\int_0^{\infty}
dp\,\log\left[1+e^{-\beta(\sqrt{p^2+M_{q}^2}\pm\mu)}\right]\;.
\label{finaloneloop}
\eqa
\end{widetext}
The full one-loop effective potential is then given by the sum of 
Eqs.~(\ref{treeb}),~(\ref{finvac}), and~(\ref{finaloneloop}).
The renormalized vacuum energy ${\cal E}_0$, background field $B$,
mass $m$, and 
quartic coupling $\lambda$ satisfy the renormalization
group equations
\bqa
\Lambda{d{\cal E}_0\over d\Lambda}&=&{m^4\lambda\over8\pi^2}\;,\\
\Lambda{d{B}^2\over d\Lambda}&=&
-{(qB)^2\over24\pi^2}
-N_c\sum_f{(q_fB)^2\over6\pi^2}
\;,\\
\Lambda{dm^2\over d\Lambda}&=&{m^2\lambda\over8\pi^2}\;,\\
\Lambda{d\lambda\over d\Lambda}&=&
{\lambda^2\over16\pi^2}-{3N_cN_fg^4\over\pi^2}\;.
\eqa
Using the renormalization-group equations, we conclude that the one-loop 
effective potential is renormalization group invariant.

We close this section by taking the limit $B\rightarrow0$
in Eqs.~(\ref{finvac}) and~(\ref{finaloneloop}).
We then need the large-$x$ behavior of 
$\zeta^{(1,0)}(-1,{1\over2}+x)$ and $\zeta^{(1,0)}(-1,x_f)$. These are given by
\bqa
\label{expand1}
\zeta^{(1,0)}(-1,{1\over2}+x)
&=&{1\over2}x^2\left(\log x-{1\over2}\right)+...\;, 
\\ \nonumber
\zeta^{(1,0)}(-1,x_f)
&=&{1\over2}x^2_f\left(\log x_f-{1\over2}\right)
-{1\over2}x_f\log x_f+...\;.
\\&&
\label{expand2}
\eqa
Inserting the expansions~(\ref{expand1}) and~(\ref{expand2})
into~(\ref{finvac}), we obtain the standard one-loop vacuum term
\bqa\nonumber
{\cal V}_{1}^{\rm vac}&=&-
{m_{\sigma}^4\over64\pi^2}\left[
\log{\Lambda^2\over m_{\sigma}^2}+{3\over2}\right]-
{3m_{\pi}^4\over64\pi^2}\left[
\log{\Lambda^2\over m_{\pi}^2}+{3\over2}\right]
\\ &&
+{N_cN_fm_{q}^4\over16\pi^2}
\left[\log{\Lambda^2\over m_q^2}+{3\over2}\right]\;.
\label{finva2}
\eqa
In Eq.~(\ref{finaloneloop}), we change variable $p_{\perp}^2=2|qB|n$, which
yields $p_{\perp}dp_{\perp}=|qB|dn$. Replacing the sum by integrals, we
obtain
\begin{widetext}
\bqa\nonumber
{\cal V}_{1}^{T,\mu}
&=& \nonumber{T\over2\pi^2}\int_0^{\infty}
dp\,p^2\log\left[1-e^{-\beta\sqrt{p^2+m_{\sigma}^2}}\right]
+{3T\over2\pi^2}\int_0^{\infty}
dp\,p^2\log\left[1-e^{-\beta\sqrt{p^2+m_{\pi}^2}}\right]
\\ && 
-
{N_cN_fT\over\pi^2}\int_0^{\infty}
dp\,p^2\log\left[1+e^{-\beta(\sqrt{p^2+m_q^2}\pm\mu)}\right]\;.
\label{finaloneloo2}
\eqa
The full one-loop effective potential at $B=0$ is the sum of
Eqs.~(\ref{treeb}), ~(\ref{finva2}) and~(\ref{finaloneloo2}).
\end{widetext} 

\section{Thermodynamics and numerical results}
In the previous section, we calculated the one-loop effective potential
at finite temperature $T$ and finite baryon chemical potential $\mu_B$. 
A common approximation in the QM model is 
to neglect the quantum and thermal fluctuations of the mesons,
which is equivalent to the large-$N_c$ limit. One 
hopes that the contributions from the quarks include the most important
effects~\cite{Scavenius:2000qd,fraga1}. 
In this paper, we will apply this approximation and defer
the use of more sophisticated methods to a subsequent paper~\cite{sub}.
The one-loop effective potential then reduces to
\begin{widetext}
\bqa\nonumber
{\cal V}_{0+1}
&=&{1\over2}m^2v^2+{\lambda\over24}v^4
-hv
+
{N_cm_{q}^4\over16\pi^2}
\sum_{f}
\left[
\log{\Lambda^2\over2|q_fB|}+1
\right]
-{N_c\over2\pi^2}\sum_f
(q_fB)^2\zeta^{(1,0)}(-1,x_f)
\\&& 
-{N_cm_{q}^2\over8\pi^2}\sum_f|q_fB|\log{m_{q}^2\over2|q_fB|}
-N_c\sum_{s,f,n}{|q_fB|T\over2\pi^2}\int_0^{\infty}
dp\,\log\left[1+e^{-\beta(\sqrt{p^2+M_{q}^2}\pm\mu)}\right]\;,
\label{finalle}
\eqa
\end{widetext}
where we for simplicity have omitted all $v$-independent terms. 
Notice that the renormalized effective potential Eq.~(\ref{finalle})
for large values of the order parameter $v$ is unbounded from below
due to the dominant term $-N_cm_q^4\log(m_q^2/2q_fB)/16\pi^2$
that arises from $\zeta^{(1,0)}(-1,x_f)$ at large $x_f$.
The unboundedness of the fermionic functional determinant is
well known for $B=0$ and arises here as well. It is due to
the minus sign in Eq.~(\ref{fpisum}).
However, it is stabilized by the bosonic contributions
in Eq.~(\ref{finva2}) for realistic values of the coupling $\lambda$,
(cf. Eqs.~(\ref{below1})--~(\ref{below2}) and the numerical values given
below). 
Generally, the reliability of perturbative (vacuum) calculations 
and the stability of the effective potential 
are nontrivial issues, see 
e. g. for a thorough discussion~\cite{stab1,stab2,stab3}.

Let us first briefly discuss magnetic catalysis of dynamical symmetry breaking,
namely the effect that chiral symmetry 
is broken dynamically for any nonzero magnetic field
when it is intact for $B=0$.
This effect is now well 
established, see e. g. ~\cite{ebert+,dcsm1,dcsm2,dcsm3,dcsm4,dcsm5,dcsm6}.
In the present case, chiral symmetry at vanishing magnetic field $B$
corresponds to choosing $m^2>0$~\footnote{$m^2>0$ corresponds to a symmetric
vacuum state at tree level. In the NJL model, symmetry breaking
is always a loop effect which takes place for $G$
smaller than a critical coupling $G_c$.}.
The minimum of the effective potential is
found by minimizing the effective potential~(\ref{finalle})
with $\mu_B=T=h=0$. In the limit $B\rightarrow0$, we know that the
expectation value $v$ is exponentially small~\cite{ebert+} 
and so we can expand
${\cal V}_{0+1}$ around $v=0$.
This yields~\footnote{We have used
$\zeta^{(1,0)}(-1,x_f)={\rm const}-x_f\log{\sqrt{2\pi}x_f}+...$
for small values of $x_f$.}
\bqa\nonumber
{d{\cal V}_{0+1}\over dv}
&\approx& 
v\left[m^2+
{N_cg^2\over8\pi^2}\sum_f|q_fB|\log{\pi m^2_q\over |q_fB|}\right]
\\
&=&0\;.
\eqa
Either $v=0$ or we obtain $v^2$
\bqa
v^2&=&{2^{2\over3}|qB|\over3g^2\pi}\exp\left[
-{8\pi^2m^2\over N_cg^2|qB|}
\right]\;.
\label{nontri}
\eqa
The nontrivial solution Eq.~(\ref{nontri}) corresponds to the local
minimum of ${\cal V}_{0+1}$.
Eq.~(\ref{nontri}) 
has the same functional form with the replacement $1/G\rightarrow m^2/g^2$
as in the NJL-model calculations of
e. g. Ref.~\cite{ebert+}. 
The reason is that we are basically evaluating
the same fermionic functional determinant although we are using a different 
ultraviolet regulator $\Lambda$.

We next discuss the determination of the parameters of the 
Lagrangian~(\ref{lagra}). In the vacuum, we have $\mu_B=0$ and 
$f_{\pi}=v$, where $f_{\pi}$ is the pion decay constant.
At tree level, the mass parameter $m^2$, the couplings $\lambda$ and $g$,
and the symmetry breaking
parameter $h$
can be expressed in terms
of observable masses of the sigma and pions, and the pion decay constant
$f_{\pi}$:
\bqa
\label{below1}
m^2&=&-{1\over2}(m_{\sigma}^2-3m_{\pi}^2)\;,\\
\lambda&=&{3(m_{\sigma}^2-m_{\pi}^2)\over f_{\pi}^2}\;,\\
g&=&{m_q\over f_{\pi}}\;,\\
h&=&f_{\pi}m^2_{\pi}\;.
\label{below2}
\eqa
We use a sigma mass of
$m_{\sigma}=800$ MeV, a constituent quark mass of $m_q=300$ MeV,
and a pion decay constant of $f_{\pi}=93$ MeV.
In the chiral limit, $h=0$, 
this yields 
$m^2=-320000$ MeV$^2$, $\lambda=222$, and $g=3.2258$.
At the physical point, we obtain
$m^2=-291018$ MeV$^2$, $\lambda=215.29$, and $g=3.2258$.
In the remainder of the paper, we set $N_c=3$ and $c=0$.

We first consider the effective potential at $T=\mu=0$. 
As noted above, we have determined the parameters of the theory
at tree level. Quantum fluctuations, i. e. loop corrections, will modify the
classical potential and it depends on the renormalization scale 
$\Lambda$. We can choose $\Lambda$ such that 
the minimum of the one-loop effective potential
in the vacuum (for  $B=0$) 
still is at $v=f_{\pi}$. This is done by requiring
\bqa
{d{\cal V}_{0+1}\over dv}\big|_{v=f_{\pi}}&=&0\;.
\eqa
This is equivalent to requiring that the one-loop self-energy 
of the
pion at zero external momentum
vanish, $\Pi_{1,\pi}(0)=0$. From Eq.~(\ref{finva2}), this yields
\bqa
\left[
\log{\Lambda^2\over g^2f_{\pi}^2}+1
\right]&=&0\;,
\eqa
whose solution is $\Lambda=181.96$ MeV. We will use this value
in the remainder of the paper.

\begin{figure}[htb]
\vspace{-4.4cm}
\setlength{\unitlength}{1mm}
\begin{picture}(100,100)
\includegraphics[width=7.8cm]{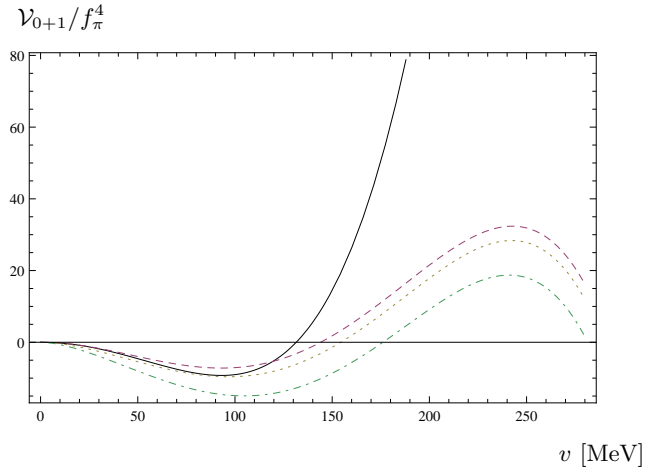}


    \put(-77,53){${\cal V}_{0+1}/f_{\pi}^4$}
    \put(-5,-5){$v$\ [MeV]}
\end{picture}
\vspace{0.4cm}
\caption{
Normalized effective potential ${\cal V}_{0+1}/f_{\pi}^4$
in the chrial limit for $T=\mu_B=0$.
Tree level (solid curve), 
one-loop with $|qB|=0$(dashed curve),
one loop with $|qB|=5m_{\pi}^2$ (dotted curve), and
one loop with $|qB|=10m_{\pi}^2$ (dash-dotted curve).
}
\label{vacpot}
\end{figure}

In Fig.~\ref{vacpot}, we plot the tree-level potential (solid curve)
as well
as the one-loop effective potential in the chiral limit
for $|qB|=0$, $|qB|=5m^2_{\pi}$, 
and $|qB|=10m^2_{\pi}$ with $T=\mu=0$ 
The dashed curve is for $|qB|=0$, the dotted curve is
$|qB|=5m_{\pi}^2$  and the dash-dotted curve is for $|qB|=10m_{\pi}^2$.
We notice that the potential becomes deeper with increasing magnetic
field and that the minimum moves to larger values of 
$v$~\footnote{Note, however, that the one-loop effective potential
for $|qB|=0$
is more shallow than the tree-level potential.}.
Thus the magnetic field enhances symmetry breaking.
This is 
in agreement with earlier 
findings~\cite{fraga1,fragapol,ebert+,dcsm1,dcsm2,dcsm3,dcsm4,dcsm5,dcsm6}.
We also note that the local maximum of the effective potential 
for large values of $v$ becomes lower with increasing magnetic field.
Thus for sufficiently strong magnetic fields, the local minimum at $v=93$ MeV
ceases to exist and the system has no longer a metastable state.
This is the instability that we discussed above.
This in disagreement with Ref.~\cite{fragapol}.
In their work, the authors subtract the fermionic vacuum fluctuations
at $B=0$. Thus they subtract a $v$-dependent contribution
given by the last term in Eq.~(\ref{finva2}) 
which leads to a one-loop effective potential 
that is independent of
the renormalization scale $\Lambda$ and is stable for large values of $v$.

\begin{figure}[htb]
\vspace{-5.1cm}
\setlength{\unitlength}{1mm}
\begin{picture}(100,100)
\includegraphics[width=6.8cm]{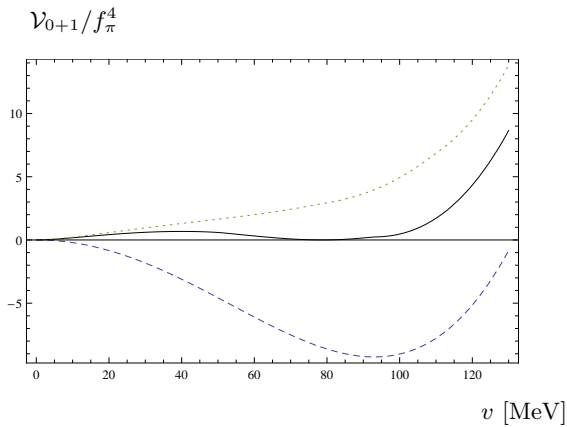}
    \put(-5,-5){$v$\ [MeV]}
    \put(-65,47){${\cal V}_{0+1}/f_{\pi}^4$}
\end{picture}
\vspace{0.6cm}
\caption{
Normalized effective potential ${\cal V}_{0+1}/f_{\pi}^4$
without vacuum fluctuations
for $T=0$ and $|qB|=5m_{\pi}^2$ as a function of $v$
for three different values of
$\mu$: $\mu=0$ (dashed curve), $\mu=\mu_c=301.6$ MeV (solid curve), 
and $\mu=358$ MeV (dotted curve).
}
\label{t0pot}
\end{figure}

It was mentioned in the introduction that the inclusion of the fermion
vacuum fluctuation term can change the order of the phase transition 
in the chiral limit as well as strongly influence physical 
observable. This was discussed in detail for $\mu_B=0$ in Ref.~\cite{bj}.
It turns out that this is the case in the entire
$T$--$\mu_B$ phase diagram for $B=0$~\cite{khan}.
We find that this is also the case in a strong magnetic field.
In Fig.~\ref{t0pot}, we show the normalized effective potential 
with no vacuum fluctuations
for three different values of the baryon chemical potential $\mu_B$
and $T=0$ as a function of $v$. 
Clearly, the transition is first order if the vacuum fluctuations
are not included.

\begin{figure}[htb]
\vspace{-4.8cm}
\setlength{\unitlength}{1mm}
\begin{picture}(100,100)
\hspace{1cm}
\includegraphics[width=6.8cm]{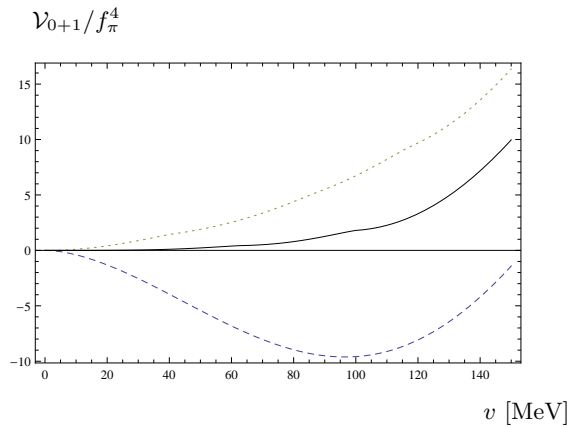}
    \put(-5,-5){$v$\ [MeV]}
    \put(-65,47){${\cal V}_{0+1}/f_{\pi}^4$}
\end{picture}
\vspace{0.6cm}
\caption{
Normalized effective potential ${\cal V}_{0+1}/f_{\pi}^4$
including vacuum fluctuations 
for $T=0$ and $|qB|=5m_{\pi}^2$ as a function of $v$
for three different values of 
$\mu$: $\mu=0$ (dashed curve), $\mu=\mu_c=324$ MeV (solid curve), 
and $\mu=380$ MeV (dotted curve).
}
\label{t0pot1}
\end{figure}

Similarly, in Fig.~\ref{t0pot1}, we show the normalized effective potential
with vacuum fluctuations
for three different values of the baryon chemical potential $\mu_B$
and $T=0$ as a function of $v$. 
The transition is second order if the vacuum fluctuations
are included.

\begin{figure}[htb]
\centering
\vspace{-5.0cm}
\setlength{\unitlength}{1mm}
\begin{picture}(100,100)
\includegraphics[width=6.8cm]{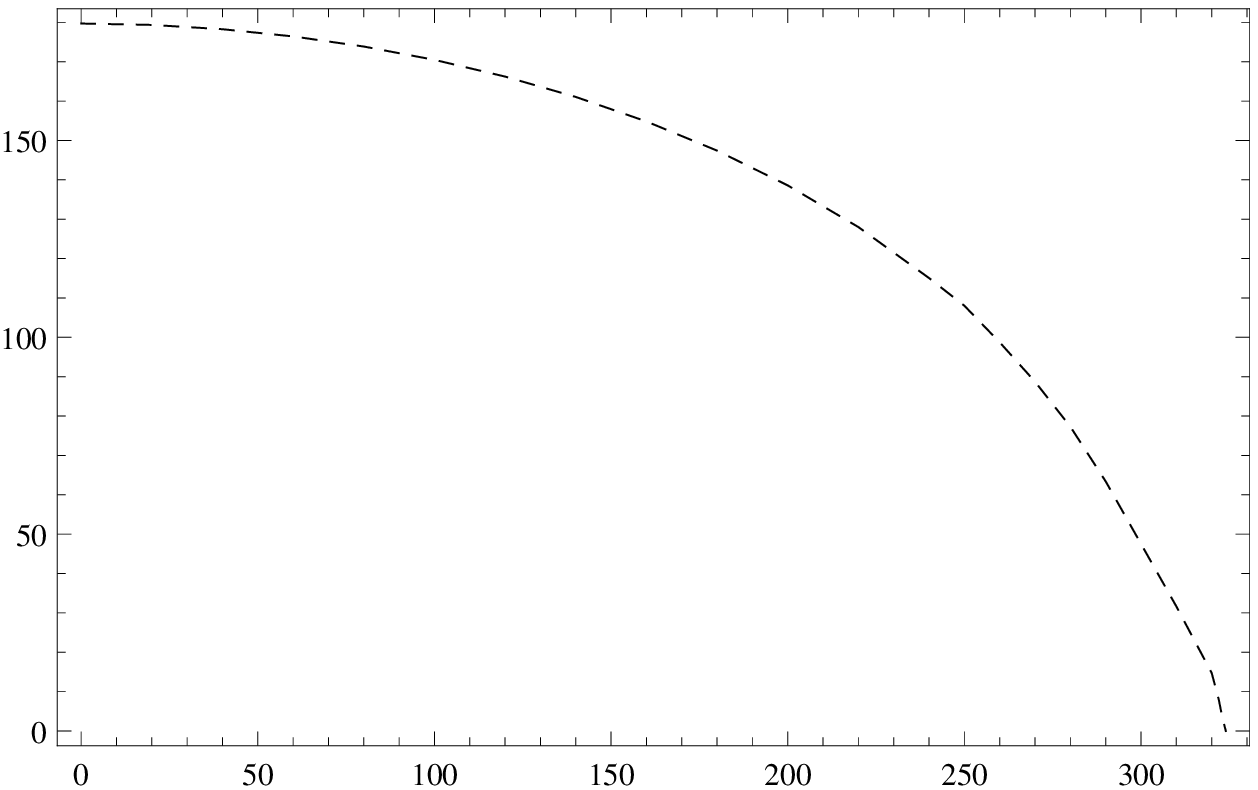}
\hspace{1cm}
    \put(-70,48){$T$\ [MeV]}
    \put(-12,-3){$\mu_B$\ [MeV]}

\end{picture}
\vspace{0.1cm}
\caption{
Phase diagram in the $\mu_B$-$T$ plane
for $|qB|=5m_{\pi}^2$ in the chiral limit. 
Fermionic vacuum fluctuations are included.
}
\label{phase1}
\end{figure}

\begin{figure}[htb]
\centering
\vspace{-5.0cm}
\setlength{\unitlength}{1mm}
\begin{picture}(100,100)
\includegraphics[width=6.8cm]{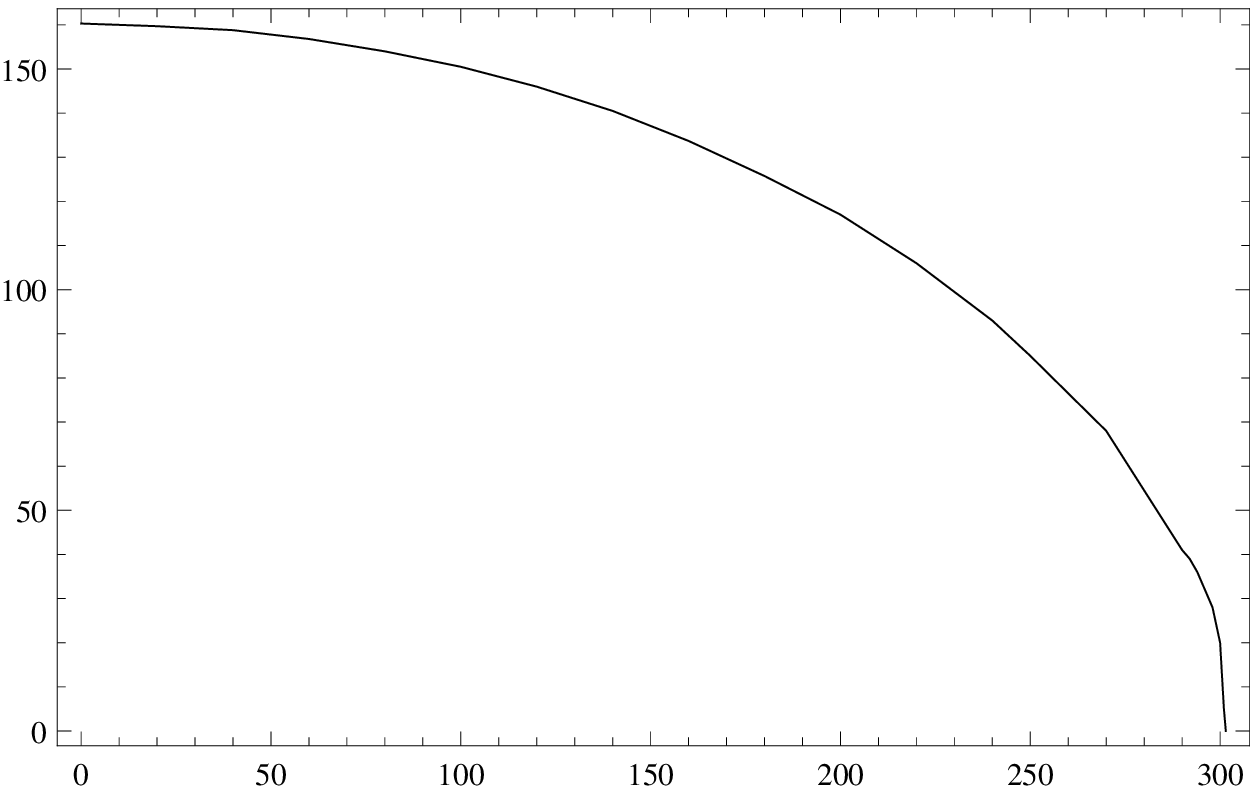}
\hspace{1cm}
    \put(-70,48){$T$\ [MeV]}
    \put(-12,-3){$\mu_B$\ [MeV]}
\end{picture}
\vspace{0.1cm}
\caption{
Phase diagram in the $\mu_B$-$T$ plane
for $|qB|=5m_{\pi}^2$ in the chiral limit. 
Fermionic vacuum fluctuations are excluded.
}
\label{phase}
\end{figure}

In Figs.~\ref{phase1}--\ref{phase}, 
we show the phase diagram as a function of $\mu_B$ and $T$
in the chiral limit. 
In Eq.~(\ref{finalle}), there is a sum over Landau levels 
and for each value of $\mu_B$, we include sufficiently many terms so that
our result for $T_c$ is converging. For example, for
$\mu_B=0$, we must typically sum the 
first ten term in the series, while for $T=0$ it suffices to include 
the first few terms in the series.

In Fig~\ref{phase1}, we have included the
fermionic vacuum fluctuations 
and the dashed curve indicates a second-order transition.
In the case where we include the quantum fluctuations, 
the critical temperature for $\mu=0$ is
$T_c=179$ MeV and the critical chemical potential for $T=0$
is $\mu_c=324$ MeV.
In comparison, the critical temperature at vanishing magnetic field
is $T_c=176$ MeV, the critical chemical potential is $\mu_c=321$ MeV, and
the  phase transition is also second order for all values of $\mu$~\cite{khan}.
Calculating the critical temperature
for $\mu_B=0$ and $|qB|=10m^2_{\pi}$ yields
$T_c=184$ MeV and so it seems to be increasing very weakly with
the strength of the magnetic field.

In Fig~\ref{phase}, we have omitted the vacuum term
and the solid curve indicates a first-order transition. 
In the case where the quantum fluctuations are omitted, 
the critical temperature for $\mu=0$ is
$T_c=160$ MeV and the critical chemical potential for $T=0$
is $\mu_c=301.6$ MeV.
In comparison, the critical temperature at vanishing magnetic field
is $T_c=179$ MeV, the critical chemical potential is $\mu_c=347$ MeV, and
the  phase transition is also first order for all values of $\mu$~\cite{khan}.
In contrast to the zero-$B$ case, the exclusion of the vacuum fluctuation
decreases the critical temperature significantly.

In Ref.~\cite{khan}, it was shown using optimized perturbation 
theory~\cite{chiku}
that the inclusion of the vacuum and thermal contributions from the
bosons significantly lowers the temperature of the transition.
Whether this implies that including the effects of the 
bosons will lower the critical temperature for nonzero $B$ as well
is not known. In particular, the $B$-dependence of the critical
temperature is an open question, see also the discussion below.

There have been a number of model calculations using 
chiral perturbation theory, the
NJL model, the NJL model coupled to the Polyakov loop, the QM model, and
the Polyakov-loop extended model. Except for chiral perturbation 
theory~\cite{chiralB},
they all predict an increase of the critical temperature with increasing
magnetic field. This is in disagreement with the lattice simulations of
Ref.~\cite{budaleik} which show a significant decrease of the crossover 
temperature. 

In Ref.~\cite{chiralB}, the authors are using chiral perturbation theory
to investigate the quark-hadron 
phase transition as a function of the magnetic field at the physical point.
They compare the pressure in the hadronic phase with that of the quark-gluon
plasma phase for an ideal gas of quarks and gluons, and subtracting
the vacuum energy due to a nonzero
gluon condensate $\langle g^2G_{\mu\nu}G^{\mu\nu}\rangle$.
For weak magnetic fields, the transition is first order. The line
of first-order transitions ends at critical point
$(\sqrt{|qB|},T)=(600,104)$ MeV. For larger values of $|qB|$, the transition
is a crossover. The authors of 
Ref.~\cite{fragapol} report that they find a first-order
transition at the physical point if they keep the remaining $B$-dependent
vacuum fluctuations and a crossover if they are ignored. 
Our sample calculations at the physical point for $|qB|=5m_{\pi}^2$ (not shown)
suggest
that the transition is a crossover (for $\mu_B=0$). This is in agreement with
the lattice simulations in Ref.~\cite{budaleik}.
The disagreement between Ref.~\cite{fragapol} and the present results
can probably 
be traced back to a different treatment of the vacuum fluctuations,
as discussed earlier.
The disagreement between various approaches deserves further investigation.

\section{Summary and Outlook}
In the present work, we have calculated the one-loop effective potential
for the quark-meson model at finite temperature and baryon density in
an external magnetic field. 
We have made the common approximation where we 
ignore all quantum and thermal effects of the bosons 
and hence treat them at tree level.
We have seen that the critical temperature and the order of the 
phase transition depend on whether one includes the vacuum fluctuations.
Taking a model seriously means including the effects of all its degrees of 
freedom. It therefore questionable to throw
out some terms unless one can show that they are not important. For example, 
the effective potential is receiving
contributions from vacuum fluctuations from all energies scales up to the 
ultraviolet cutoff $\Lambda$ of the effective theory.
This is the region of validity of the low-energy effective theory and it is 
therefore not obvious that they can be
omitted.

Given the conflicting results of various model calculations and lattice 
calculations, clearly more work needs to be done. On the analytical
side, an important extension is to include the vacuum and thermal
effects of the bosons.
The dominant contributions at weak coupling from the bosons arise from the 
daisy or ring diagrams. There has been some work on resumming the ring
diagrams in the presence of a 
magnetic field. For example, 
the authors of Ref.~\cite{skalozub} investigated 
their role in the electroweak phase transition in the standard model.
Similarly, their effects at weak fields have been studied in 
Ref.~\cite{raya} in the context of the chiral transition. 
One
should apply sophisticated resummation techniques 
or nonperturbative methods such as 
optimized perturbation theory~\cite{chiku}, the 2PI effective action
formalism~\cite{cornwall}, or the functional renormalization 
group~\cite{wetterich}.
These methods are known to correctly predict a second-order
phase transition for $\mu_B=B=0$~\cite{chiku2,aleo,erg}.

\section*{Acknowledgments}
The authors would like to thank E.  S. Fraga for useful discussions.
R. Khan was supported by the Higher Education Commision of Pakistan (HEC).

\end{document}